\documentclass[]{aa}  
\usepackage[utf8]{inputenc}
\usepackage{graphicx}
\usepackage{amsmath}
\usepackage{txfonts}
\usepackage{fourier}
\usepackage{url}
\usepackage{longtable}
\usepackage{tikz}
\usepackage{chemformula}
\usepackage[breaklinks, colorlinks, citecolor=blue, linkcolor=blue]{hyperref}
\usepackage[normalem]{ulem}
\usepackage{fancyhdr}
\usepackage{color}

\usepackage{acronym}
\acrodef{LBT}[LBT]{Large Binocular Telescope}
\acrodef{CCF}[CCF]{Cross-Correlation Function}
\acrodef{UHJ}[UHJ]{ultra-hot Jupiter}
\acrodef{CD}[CD]{cross-desperser}

\begin{document}

\newcommand{\teff}{T$_{\rm eff}$}
\newcommand{\logg}{$\log${(g)}}
\newcommand{\met}{$[$Fe/H$]$}
\newcommand{\masc}{MASCARA-1}
\newcommand{\mascb}{MASCARA-1~b}
\newcommand{\cheops}{CHEOPS}
\newcommand{\tess}{TESS}
\newcommand{\uvis}{UVIS}
\newcommand{\aas}{AAS}
\newcommand{\vsini}{\ensuremath{v \sin i_\star}\xspace}
\newcommand{\vmic}{$V_{\rm mic}$}
\newcommand{\vmac}{$V_{\rm mac}$}
\newcommand{\kms}{km\,s$^{-1}$}
\newcommand{\otwo}{\ion{O}{2}}
\newcommand{\vsys}{$v_{\rm sys}$}
\newcommand{\kp}{$K_{\rm p}$}

\graphicspath{{figures/}}
\let\orgautoref\autoref
\makeatletter
\renewcommand*\aa@pageof{, page \thepage{} of \pageref*{LastPage}}
\def\instrefs#1{{\def\scsep{\def\scsep{,}}\@for\w:=#1\do{\scsep\ref{inst:\w}}}}
\makeatother

    \title{The PEPSI Exoplanet Transit Survey.}

   \subtitle{III: The detection of \ion{Fe}{i}, \ion{Cr}{i} and \ion{Ti}{i} in the atmosphere of \mascb\ through high-resolution emission spectroscopy.}


   \author{
G. Scandariato\inst{1} $^{\href{https://orcid.org/0000-0003-2029-0626}{\includegraphics[scale=0.5]{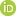}}}$,
F. Borsa\inst{2},
A.S. Bonomo,\inst{3}
B.S. Gaudi,\inst{4}
Th. Henning,\inst{5}
I. Ilyin\inst{6} $^{\href{https://orcid.org/0000-0002-0551-046X}{\includegraphics[scale=0.5]{figures/orcid.jpg}}}$,
M.C. Johnson\inst{4} $^{\href{https://orcid.org/0000-0002-5099-8185}{\includegraphics[scale=0.5]{figures/orcid.jpg}}}$,
L. Malavolta,\inst{7,8}
M. Mallonn,\inst{6}
K. Molaverdikhani\inst{9,10,5} $^{\href{https://orcid.org/0000-0002-0502-0428}{\includegraphics[scale=0.5]{figures/orcid.jpg}}}$,
V. Nascimbeni,\inst{7}
J. Patience,\inst{11}
L. Pino,\inst{12}
K. Poppenhaeger,\inst{6,13}
E. Schlawin,\inst{14}
E.L. Shkolnik,\inst{11}  $^{\href{https://orcid.org/0000-0002-7260-5821}{\includegraphics[scale=0.5]{figures/orcid.jpg}}}$,
D. Sicilia\inst{1}  $^{\href{https://orcid.org/0000-0001-7851-5168}{\includegraphics[scale=0.5]{figures/orcid.jpg}}}$,
A. Sozzetti,\inst{3}
K.G. Strassmeier,\inst{6,13}  $^{\href{https://orcid.org/0000-0002-6192-6494}{\includegraphics[scale=0.5]{figures/orcid.jpg}}}$,
C. Veillet,\inst{15}
J. Wang,\inst{4}
F. Yan\inst{16,5,17} $^{\href{https://orcid.org/0000-0001-9585-9034}{\includegraphics[scale=0.5]{figures/orcid.jpg}}}$
}

   \institute{
\label{inst:1} INAF, Osservatorio Astrofisico di Catania, Via S. Sofia 78, 95123 Catania, Italy \and
\label{inst:2} INAF - Osservatorio Astronomico di Brera, Via E. Bianchi, 46, 23807 Merate (LC), Italy \and
\label{inst:3} INAF – Osservatorio Astrofisico di Torino, Strada Osservatorio 20, I-10025 Pino Torinese, Italy \and
\label{inst:4} Department of Astronomy, The Ohio State University, 4055 McPherson Laboratory, 140 West 18th Ave., Columbus, OH 43210 USA \and
\label{inst:5} Max-Planck-Institut f\"ur Astronomie, K\"onigstuhl 17, D-69117 Heidelberg, Germany \and
\label{inst:6} Leibniz-Institute for Astrophysics Potsdam (AIP), An der Sternwarte 16, D-14482 Potsdam, Germany \and
\label{inst:7} INAF – Osservatorio Astronomico di Padova, Vicolo dell’Osservatorio 5, I-35122 Padova, Italy \and
\label{inst:8} Dipartimento di Fisica e Astronomia “Galileo Galilei”, Universita degli Studi di Padova, I-35122 Padova, Italy \and
\label{inst:9} Universit\"ats-Sternwarte, Ludwig-Maximilians-Universit\"at M\"unchen, Scheinerstrasse 1, D-81679 M\"unchen, Germany \and
\label{inst:10} Exzellenzcluster Origins, Boltzmannstra{\ss}e 2, 85748 Garching, Germany \and
\label{inst:11} School of Earth and Space Exploration, Arizona State University, 660 S. Mill Ave., Tempe, Arizona 85281, USA \and
\label{inst:12} INAF – Osservatorio Astrofisico di Arcetri, Largo Enrico Fermi 5, 50125 Firenze \and
\label{inst:13} Institute of Physics \& Astronomy, University of Potsdam, Karl-Liebknecht-Str. 24/25, D-14476 Potsdam, Germany \and
\label{inst:14} Steward Observatory, University of Arizona, 933 N. Cherry Ave., Tucson, AZ 85721, USA \and
\label{inst:15} Large Binocular Telescope Observatory, 933 N. Cherry Ave., Tucson, AZ 85721, USA \and
\label{inst:16} Department of Astronomy, University of Science and Technology of China, Hefei 230026, China \and
\label{inst:17} Institute of Astrophysics, University of G\"ottingen, Friedrich-Hund-Platz 1 , D-37077 G\"ottingen, Germany
}

   \date{accepted by A\&A April 6, 2023}

 
  \abstract{Hot giant planets like \mascb\ are expected to have thermally inverted atmospheres, that makes them perfect laboratory for the atmospheric characterization through high-resolution spectroscopy. Nonetheless, previous attempts of detecting the atmosphere of \mascb\ in transmission have led to negative results.}
  {In this paper we aim at the detection of the optical emission spectrum of \mascb.}
  {We used the high-resolution spectrograph PEPSI to observe \masc\ (spectral type A8) near the secondary eclipse of the planet. We cross-correlated the spectra with synthetic templates computed for several atomic and molecular species.}
  {We obtained the detection of \ion{Fe}{i}, \ion{Cr}{i} and \ion{Ti}{i} in the atmosphere of \mascb\ with a S/N $\approx$7, 4 and 5 respectively, and confirmed the expected systemic velocity of $\approx$13 km/s and the radial velocity semi-amplitude of \mascb\ of $\approx$200 km/s. The detection of Ti is of particular importance in the context of the recently proposed Ti cold-trapping below a certain planetary equilibrium temperature.}
  {We confirm the presence of an the atmosphere around \mascb\ through emission spectroscopy. We conclude that the atmospheric non detection in transmission spectroscopy is due to the high gravity of the planet and/or to the overlap between the planetary track and its Doppler shadow.}


\authorrunning{Scandariato et al.}

   \maketitle
%

\section{Introduction}

Atmospheric studies of \acp{UHJ}, that is short-period giant planets with equilibrium temperatures exceeding $\sim$2000 K because of the strong incident stellar flux, have been experiencing a strong increase in the last years.
High-resolution transmission spectroscopy of \acp{UHJ}, which has been proven to be the most powerful technique to investigate their atmospheres up to now, has revealed the presence of many atomic species in their atmospheres  \citep[e.g.,][]{Yan2018,hoeijmakers19,casasayas20,borsa21,borsaNatAs,tabernero21,kesseli22,prinoth2022} and started probing their dynamics \citep[e.g.,][]{ehrenreich20,seidel21,borsaNatAs,kesseli2021}.

\object{\mascb}\ \citep{Talens2017} is one of the hottest known \acp{UHJ} orbiting one of the brightest transiting exoplanet host stars (see Table~\ref{tab:parameters} for some details of the system parameters). Nevertheless, high-resolution transmission spectroscopy studies failed to find any trace of its atmosphere \citep{Stangret2022,Casasayas2022}. This has been attributed to the possibility that the atmospheric scale height is too small for successful transmission spectroscopy mainly because of the large surface gravity \citep[$\log g_p=3.6$ (cgs, )][]{Casasayas2022}. Moreover, due to the geometric configuration of the system, the planetary track \citep[which depends on the planetary velocity in the stellar rest-frame and the projected spin-orbit angle, ][]{Casasayas2022} almost completely overlaps with the Doppler shadow (which depends on the stellar rotation).

A complementary technique to transmission spectroscopy, that probes the terminator region of the planet, is phase-resolved emission spectroscopy, that searches for emission signals from the planetary dayside. This technique exploits the fact that, for \acp{UHJ}, the atmospheric temperature-pressure profile is expected to be inverted \citep[e.g.,][]{Lothringer2019}, causing atomic lines to be seen in emission in the dayside of the planet. Since with this technique we do not need to see through the atmosphere, the atmospheric density issue can be overcome. This technique has already shown that it can be as useful as transmission spectroscopy to give clues about the atmospheres of \acp{UHJ} \citep[e.g.,][]{Pino2020,nugroho2020,borsa2022}, and is possibly the most adapt to investigate the atmosphere of \mascb\ \citep{Casasayas2022}.

This technique has already shown to be complementary to transmission spectroscopy in probing the atmospheres (dayside and terminator region respectively) of \acp{UHJ} \citep[e.g.,][]{Pino2020,nugroho2020,borsa2022}. Furthermore, emission spectroscopy is likely the most adapt to investigate the atmosphere of \mascb\ \citep{Casasayas2022}.

In this work we analyze high resolution spectra of \object{\masc}, for which \citet{Holmberg2022} already detected H$_2$O and CO in emission in the near infrared. We used the PEPSI spectrograph to observe \mascb\ when it shows its dayside, looking for signals of the planetary atmosphere in the optical and near infrared bands. In Sect.~\ref{sec:observations} we present the data we collected and their reduction. In Sect.~\ref{sec:extraction} we describe how we extracted the planetary signal and we conclude in Sect.~\ref{sec:conclusions} with the final remarks.


\section{Observations and data reduction}\label{sec:observations}

The PEPSI Exoplanet Transit Survey (PETS) is a collaboration which aims at the spectroscopic analysis of exoplanetary transits and secondary eclipses, together with the characterization of their host stars \citep{Keles2022,Johnson2022}. The project takes advantage of the light collecting power of the \ac{LBT} and the high-resolution spectroscopy of the Potsdam Echelle Polarimetric and Spectroscopic Instrument \citep[PEPSI, ][]{Strassmeier2015,Strassmeier2018}.

In the framework of the PETS programs, we observed the \masc\ (HD~201585) system, consisting of a Jupiter-sized planet orbiting a fast rotating (\vsini=$109\pm4$~km/s) A8 type star every $\sim$2.14 days \citep{Talens2017}. Given the architecture of the system, \mascb\ is one of the most irradiated exoplanets known to date. \citet{Hooton2022} have recently re-analyzed the system confirming the orbital misalignment and estimating an extremely high dayside temperature larger than 3000~K. The stellar, planetary and orbital parameters most relevant for our work are listed in Table~\ref{tab:parameters}.

\begin{table*}
\caption{Stellar and system parameters.}\label{tab:parameters}      
\centering          
\begin{tabular}{l c c c r}     
\hline\hline       
Parameter & Symbol & Units & Value & Ref.\\
\hline
Stellar effective temperature & $\rm T_{eff}$ & K & $7490\pm150$ & \citet{Hooton2022}\\
Stellar radius & $\rm R_*$ & $\rm R_\sun$ & 2.1(2) & \citet{Talens2017}\\
Stellar mass & $\rm M_*$ & M$_{\rm\sun}$ & 1.90(7) & \citet{Talens2017}\\
Stellar age & $t_*$ & Gyr & 1.0(2) & \citet{Talens2017} \\
Orbital period & $P_{\rm orb}$ & days & 2.1487738(8) & \citet{Hooton2022}\\
Time of transit & $T_{\rm 0}$ & $\rm BJD_{TDB}$ & 2458833.48815(9) & \citet{Hooton2022}\\
Systemic velocity & \vsys & km/s & 11.20(8) & \citet{Talens2017}\\
RV semi-amplitude of the star & $\rm K_{*}$ & m/s & $400\pm100$ & \citet{Talens2017}\\
Planet mass & $\rm M_p$ & $\rm M_{Jup}$ & 3.7(9) & \citet{Talens2017}\\
Planet radius & $\rm R_p$ & $\rm R_{Jup}$ & 1.5(3) & \citet{Talens2017}\\
RV semi-amplitude of the planet & $\rm K_{\rm p}$ & km/s & $204.4\pm2.5$ & derived from $P_{\rm orb}$ and $\rm M_*$ \\
\hline                  
\end{tabular}
\end{table*}

Aiming at the emission spectrum of the planet, we collected the spectra of the system on two nights, at orbital phases before (June,10 2021, N1 hereafter) and after (June,19 2021, N2 hereafter) secondary eclipse, when the planet-to-star flux ratio is most favorable for the planetary emission spectroscopy. For each night of observations, we used the \ac{LBT} in binocular mode, using  the \ac{CD} of the PEPSI spectograph CD3 in the blue arm (4800-5441~\AA, R$\sim$115\,000) and CD6 in the red arm (7419-9067~\AA, R$\sim$115\,000). The exposure time of each camera was 10 minutes and, despite a negligible lag of a few seconds due to different duty cycles in the two arms, we can assume perfect simultaneity between the series of spectra of the blue and red arms. Where not specified, in the following we will always refer to a single spectrum observed by \ac{LBT}, covering both the CD3 and CD6 spectral ranges. After some preliminary analysis, we rejected the last four spectra taken on N1 that were obtained during twilight and one spectrum on N2 with S/N<100. The log-book of the remaining observations is reported in Table~\ref{tab:logbook}.

\begin{table*}
\caption{Log-book of the observations.}\label{tab:logbook}      
\centering          
\begin{tabular}{c c c c c c c c}     
\hline\hline       
Night & Date & Time\tablefootmark{a} [UT] & Exp. time [min] & S/N$_{\rm CD3}$ & S/N$_{\rm CD6}$ & Airmass & Orbital phase\tablefootmark{b}\\
\hline
N1 & 2021-06-10 & 07:12:04.6--10:36:56.1 & 10.0 & 263--380 & 300--380 & 1.9162--1.0860 & 0.3867--0.4530\\
N2 & 2021-06-19 & 06:41:43.2--11:00:47.6 & 10.0 & 225--334 & 279--374 & 1.8608--1.0826 & 0.5650--0.64947\\
\hline
\end{tabular}
\tablefoot{
    \tablefoottext{a}{Start time of the spectra taken with CD3. The corresponding start time of the CD6 spectra drift by less than 5 seconds along the series.}
    \tablefoottext{b}{Computed using the ephemeris in Table~\ref{tab:parameters}. For reference, the first and fourth contact during the eclipse happen at orbital phases 0.4590 and 0.5410 respectively.}
}
\end{table*}

Data reduction is performed in the same way as presented in \citet{Keles2022}. In summary, all science and calibration images after bias subtraction were first corrected for the master flat field image to eliminate CCD pixel-to-pixel noise. The optimal extraction of each of the five slices of all spectral orders is done by fitting the smoothed spatial profile to the raw data. All extracted slices are then averaged with weights of their pixels variances to the wavelength grid of the middle slice. The continuum fit was performed with a constrained 2D polynomial to co-align the overlapping parts of the spectral orders after continuum normalization. Finally, the spectral orders were rectified into a single 1D spectrum with weighted average of overlapping parts of orders while keeping the wavelength scale of the preceding orders. The extracted 1D spectra from the two LBT mirrors are averaged  with weights and subjected to a final continuum correction with a smoothing spline in clip-and-fit mode which selects only the continuum wavelength pixels. The wavelength scale of the resulting spectra are then transformed into the solar system barycenter by using JPL ephemeris. In this work we perform our analysis on the 1D final spectra.

\begin{figure}
    \centering
    \includegraphics[width=\linewidth]{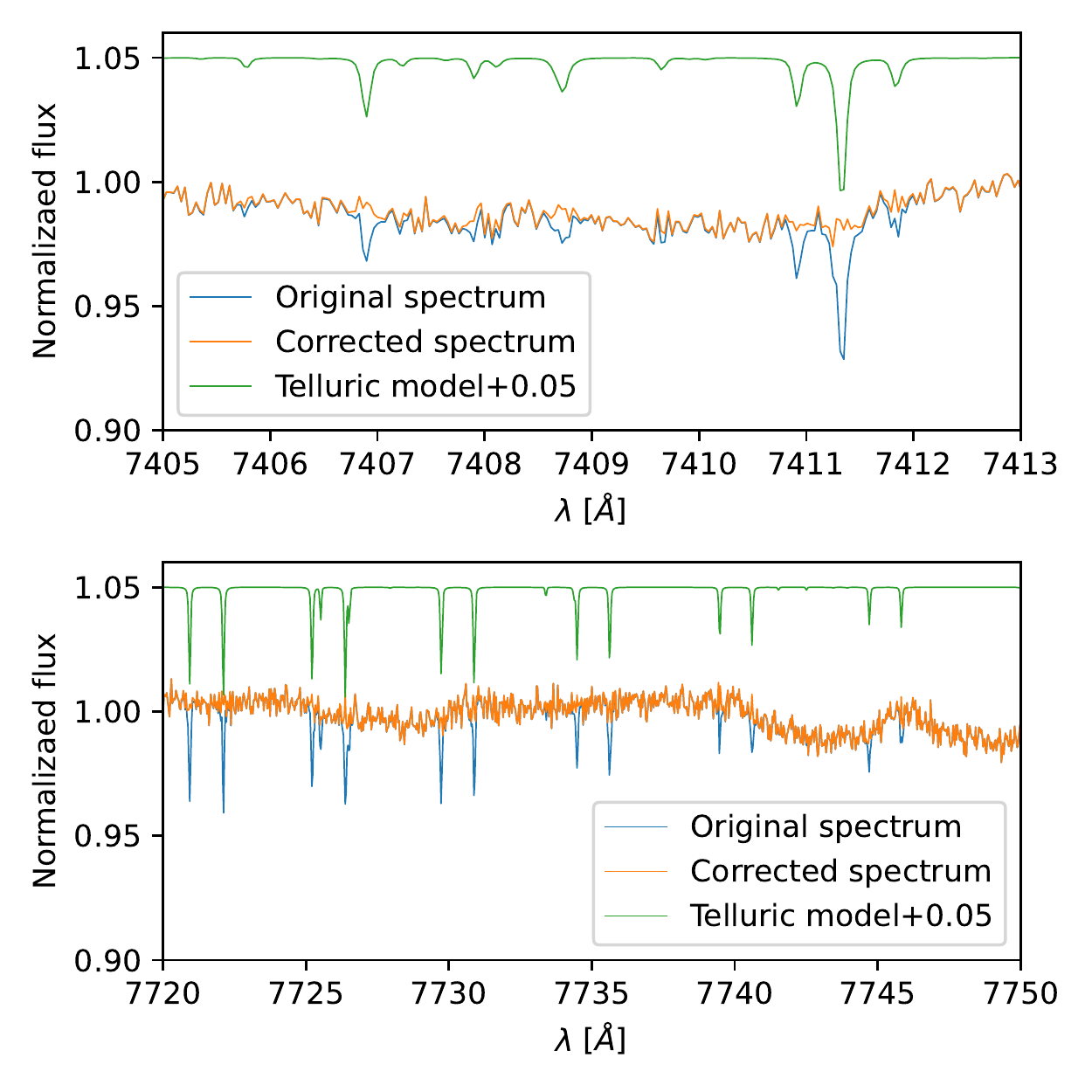}
    \caption{Example of the telluric correction performed using MOLECFIT, showing the case of the H$_2$O (top panel) and O$_2$ (bottom panel) telluric lines. In each panel, the telluric spectrum has been shifted upwards for clarity.}
    \label{fig:telluric}
\end{figure}

To correct the telluric absorption in the spectra, we computed the best telluric model including both $\rm H_2O$ and $\rm O_2$. To this purpose, we injected to MOLECFIT \citep{Kausch2015,Smette2015} the spectral ranges least affected by the stellar absorption lines
listed in Table~\ref{tab:telluric}. We then used the best-fit telluric model to correct the full spectra. An example of the telluric correction is shown in Fig.~\ref{fig:telluric}.

\begin{table}
\caption{List of the wavelength ranges used to compute the telluric absorption model.}\label{tab:telluric}      
\centering          
\begin{tabular}{c}     
\hline\hline       
Wavelength range [\AA]\\
\hline
5030--5045\\
5045--5055\\
5055--5065\\
5065--5080\\
7390--7400\\
7405--7415\\
7720--7740\\
7950--7990\\
8005--8030\\
8400--8440\\
8560--8575\\
8890--8905\\
8095--8130\\
8358--8380\\
\hline                  
\end{tabular}
\end{table}

We then shifted the spectra to the stellar rest frame using the system parameters in Table~\ref{tab:parameters} and resampled them to a common wavelength step of 0.01~\AA\ for CD3 and 0.03~\AA\ for CD6. We computed the median of the spectra, thus obtaining a high S/N representation of the stellar spectrum void of the contribution of the emission spectrum of the planet. As a matter of fact, the planetary emission lines are expected to shift due to the Doppler effect from one observation to the next one in the stellar rest frame, and get canceled out by the median computation  \citep[see][for a detailed description]{Scandariato2021}.

We thus divided each spectrum by the average spectrum and analyzed the residuals of each spectrum with respect to median spectrum. This analysis emphasized the presence of a few spikes, mainly due to cosmic rays hits, which we clipped out with a 5$\sigma$ rejection criterion and substituted with the median flux. Secondly, we identified and masked out the spectral regions where the telluric correction left residuals stronger than the spectral noise, typically in the case of strong or saturated telluric O$_2$ lines in the red part of the spectra (7590--7700~\AA, 8100--8370~\AA, $>$8900~\AA). Finally, we refined the normalization of each spectrum by fitting the residuals with a low order Savitzky-Golay filter, which preserves the spectral lines while correcting second order inaccuracies in the reduction and normalization of the spectra.

\section{Method}\label{sec:extraction}

We divided the each spectrum in the N1 and N2 series by the corresponding median spectrum: this procedure removed the stellar absorption lines while preserving the planetary emission spectrum. To all effects, these spectra represent a phase-resolved series of continuum-normalized planetary emission spectra.

To extract the planetary signal buried in the noise we used the \ac{CCF} technique \citep{Snellen2010}, which has already been successfully applied for the analysis of emission spectroscpy \citep[see for example][]{nugroho2020,Yan2020,Yan2022}. Given a spectrum $F(\lambda)$ and a template $T(\lambda,v)$ Doppler-shifted at the velocity $v$, the \ac{CCF} $C(v)$ is computed as
\begin{equation}
    C(v)=\Sigma_{i=0}^N F(\lambda_i)T_i(\lambda_i,v),
\end{equation}
where the summation is extended over all the $N$ spectral points in the observed spectrum. In principle, the contribution of the echelle orders can be weighted by their respective S/N. In practice, since the S/N does not change significantly along the orders, for the sake of simplicity we prefer not to weight by the S/N. This also preserves the relative contribution of the spectral lines in the templates.

We generated emission template spectra using \texttt{petitRADTRANS} \citep{molliere2019}, assuming solar metallicity and equilibrium chemistry. We use the T-P profile presented in \citet{Lothringer2019} for a planet with $T_{\rm eq}$=2250~K orbiting an A0 host star, with no TiO in the atmosphere (see their Fig.~3). The choice of using the profile with no TiO is motivated by the fact that so far it has been detected only in the atmosphere of WASP-189~b \citep{prinoth2022} and WASP-33~b \citep{Nugroho2017,cont2021}, while its detection is at least controversial for other \acp{UHJ}, despite the predictions to be in their atmospheres and being a possible cause of the thermal inversion \citep[e.g.,][]{hubeny2003,fortney2008}. The absence of TiO is also supported a posteriori by the fact that we do not detect it in the collected spectra (TiO was modeled and searched for also with the theoretical T-P profile assuming TiO in the atmosphere). To scale the templates to the same units as the reduced spectra, we added the stellar blackbody spectrum with T$_{\rm eff}$=7490~K (Table~\ref{tab:parameters}) and continuum normalized them.

We generated the templates for the species \ion{Fe}{i}, \ion{Fe}{ii}, FeH, \ion{Ca}{i}, CaH, \ion{Cr}{i}, \ion{Si}{i}, \ion{Ti}{i}, \ion{V}{I}, \ion{Y}{i}, TiO, VO, which have the largest number of lines in the wavelength range covered by our spectra \citep[see for example Appendix A.1 in ][]{Kitzmann2021}.
We removed all the spectral lines in the templates that are weaker than 1\% of the strongest line in the template, as they are prone to introduce noise in the CCF computation.

\begin{figure}
    \centering
    \includegraphics[width=\linewidth]{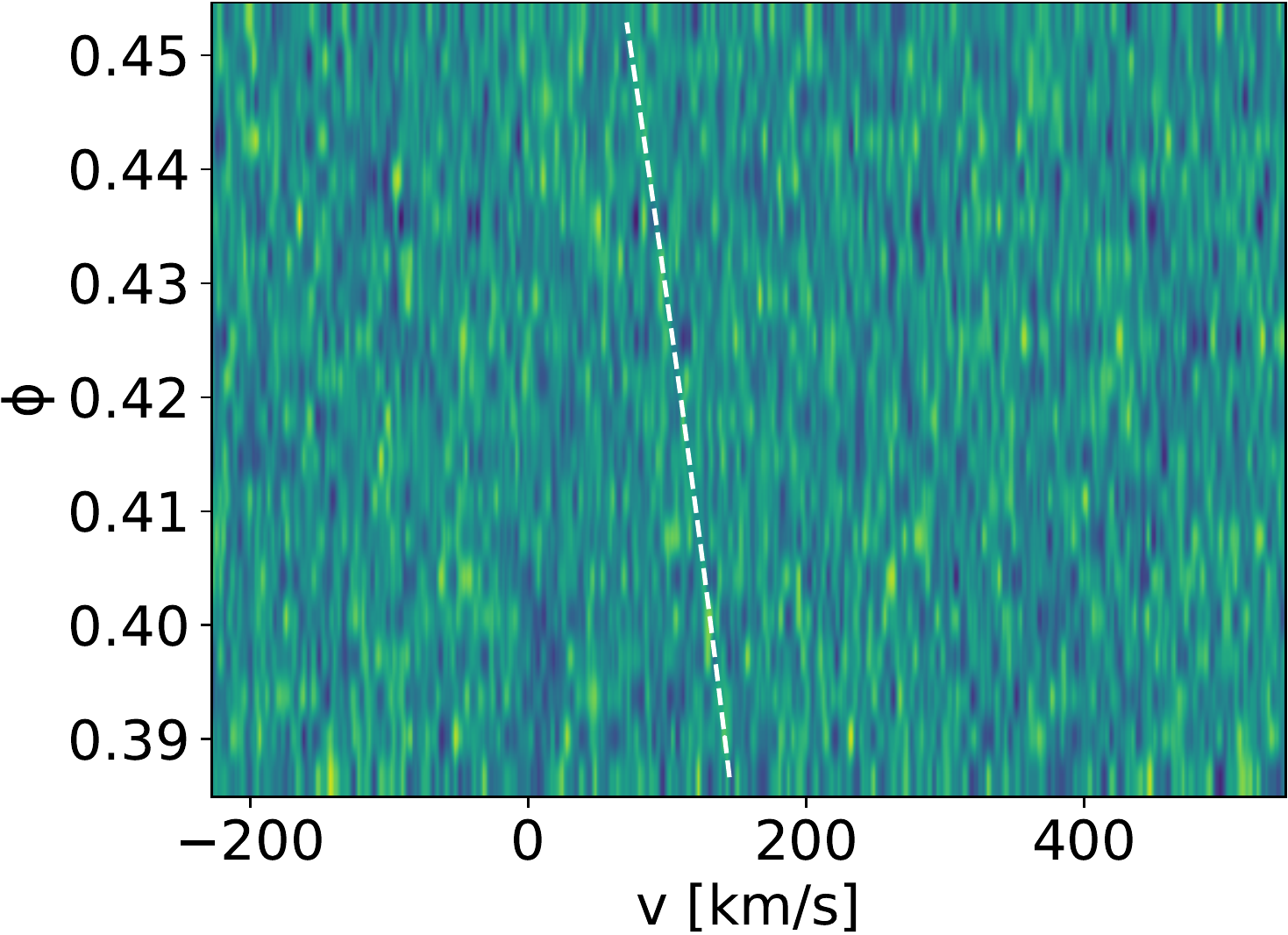}\\
    \includegraphics[width=\linewidth]{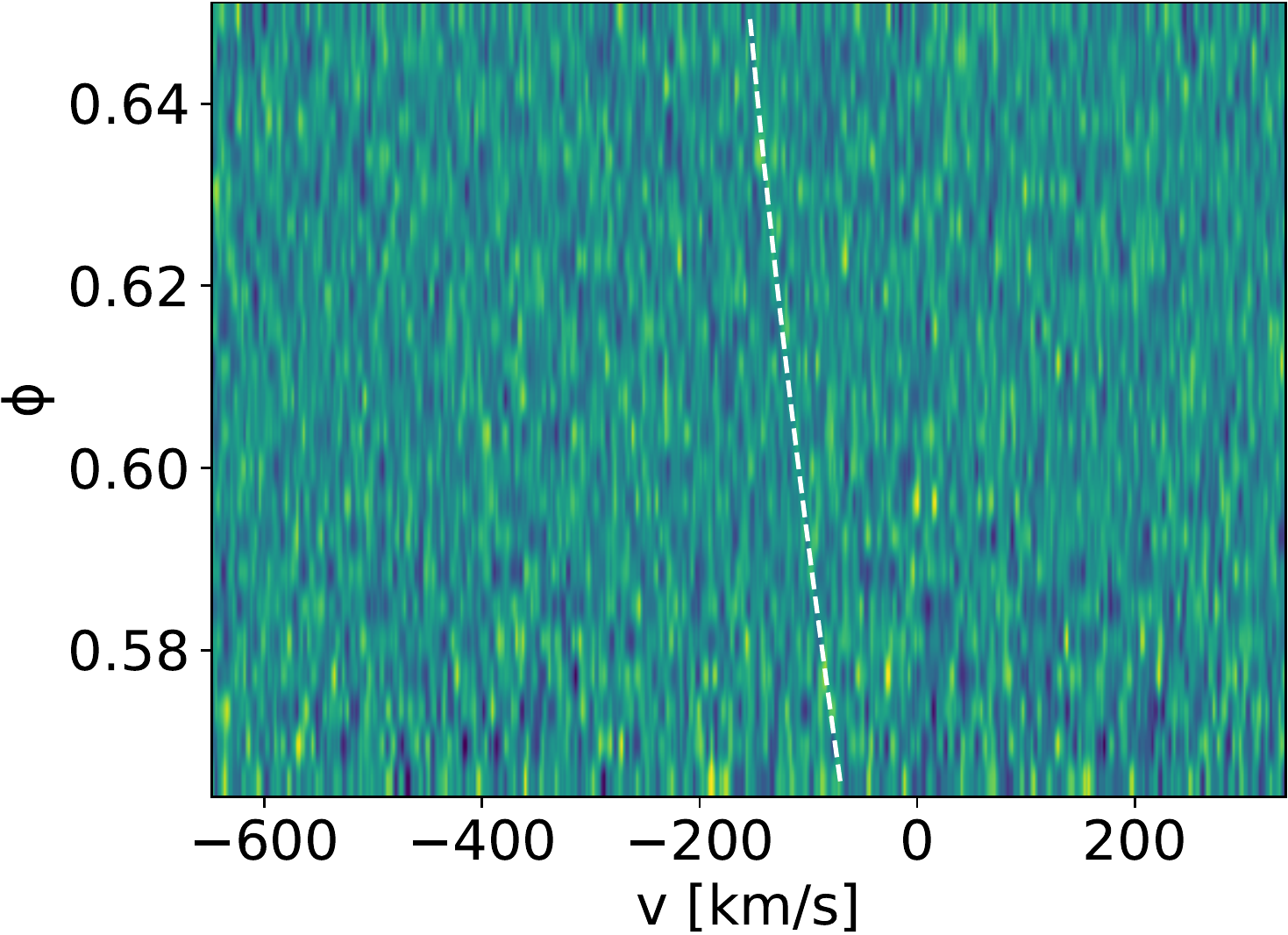}
    \caption{\acp{CCF} of the planetary emission spectrum for night N1 (top panel) and N2 (bottom panel) as a function of the planetary orbital phase. The \acp{CCF} shown in these plots are computed using the \ion{Fe}{i} spectral template, for which we claim the strongest detection (see Table~\ref{tab:peaks}). In each panel, the white dashed line marks where the planetary emission signal is expected according to the system parameters in Table~\ref{tab:parameters}. The emission trace is not obvious in this 2D plot and is only detected significantly in the \kp-\vsys\ map as shown in Fig.~\ref{fig:kpvsysFeI}.}
    \label{fig:ccfStack}
\end{figure}

We used the templates to compute the \ac{CCF} of each planetary emission spectrum. For each night and for each template we thus derived a stack of \acp{CCF}. In a 2D representation, the planetary signal is expected to show up as an emission feature moving in the velocity space according to the planetary keplerian motion in the stellar rest frame. This approach did not lead us to detection because the expected signal, if present, is smaller than the noise of the \acp{CCF} (see for example Fig.~\ref{fig:ccfStack}).

To enhance the detectability of the planetary signal it is a common practice to re-align and stack the \acp{CCF} into the planetary rest frame. To this purpose, we iteratively assumed a value for \vsys\ in the range -5--30~km/s with a step of 0.5~km/s and a value for the planetary radial velocity semi-amplitude \kp\ in the range 50--300~km/s with a step of 1~km/s. For each couple of \vsys\ and \kp\ we thus re-centered each \ac{CCF} into the planetary rest-frame, assuming the ephemeris in Table~\ref{tab:parameters}, and we computed the average planetary \ac{CCF} at velocity $v=0$, where the peak of the planetary signal is expected. This procedure generated a 2D \kp--\vsys\ map that is expected to show a peak at the (\kp,\vsys) coordinates coinciding with the \kp\ and \vsys\ derived for the planet and the system. Conversely, pixels far off the expected peak should be randomly distributed around zero depending on the noise of the spectra and how it interferes with the templates used for the cross-correlation. To estimate the S/N of the detection, we normalized the \kp--\vsys\ maps by the mean absolute deviation of the pixels in the map. The computation is extended to the whole map and, by construction, is robust against outlier pixel values.

\begin{figure*}
    \centering
    \includegraphics[width=.3\linewidth]{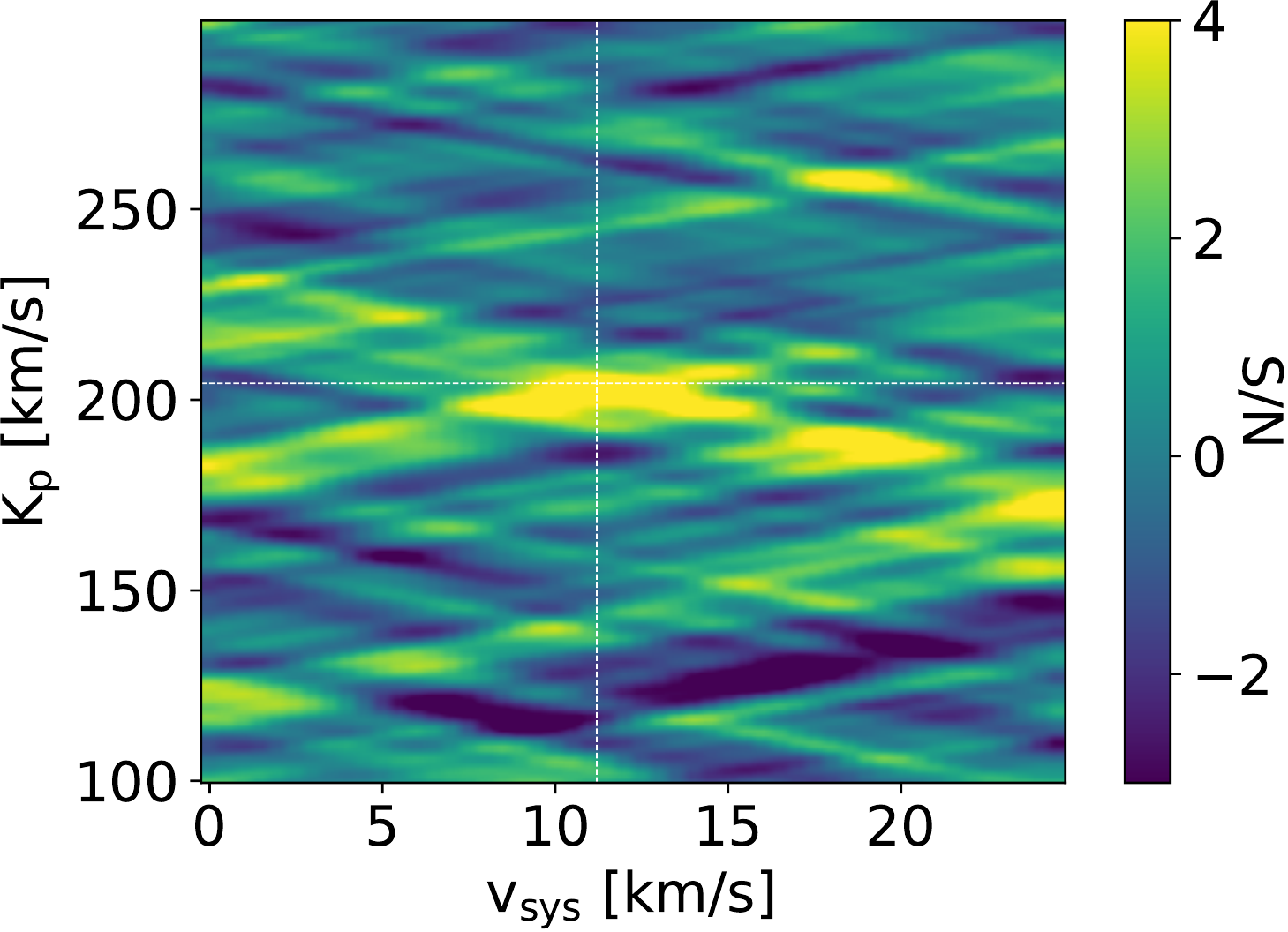}\hspace{.5cm}
    \includegraphics[width=.3\linewidth]{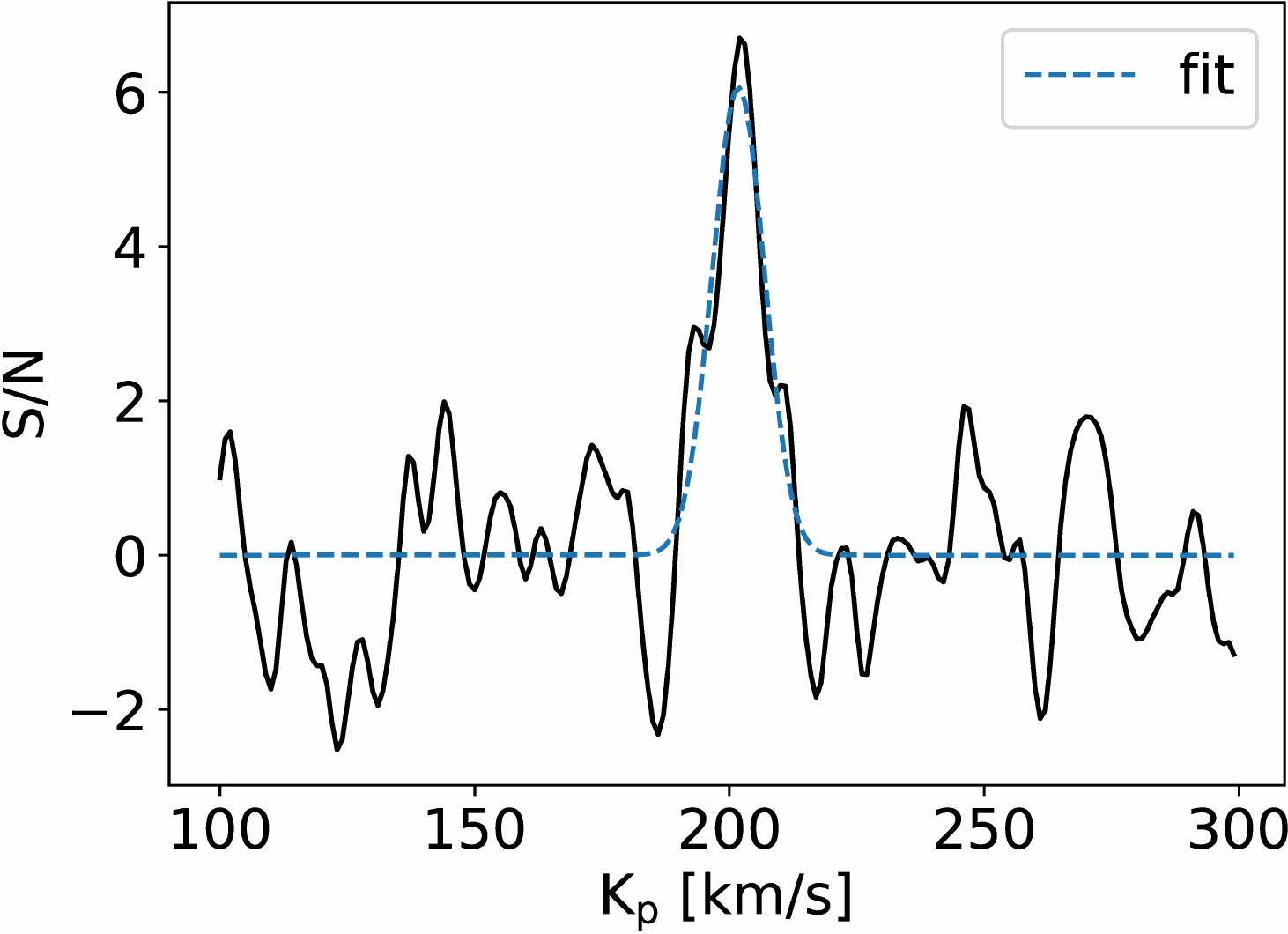}\hspace{.5cm}
    \includegraphics[width=.3\linewidth]{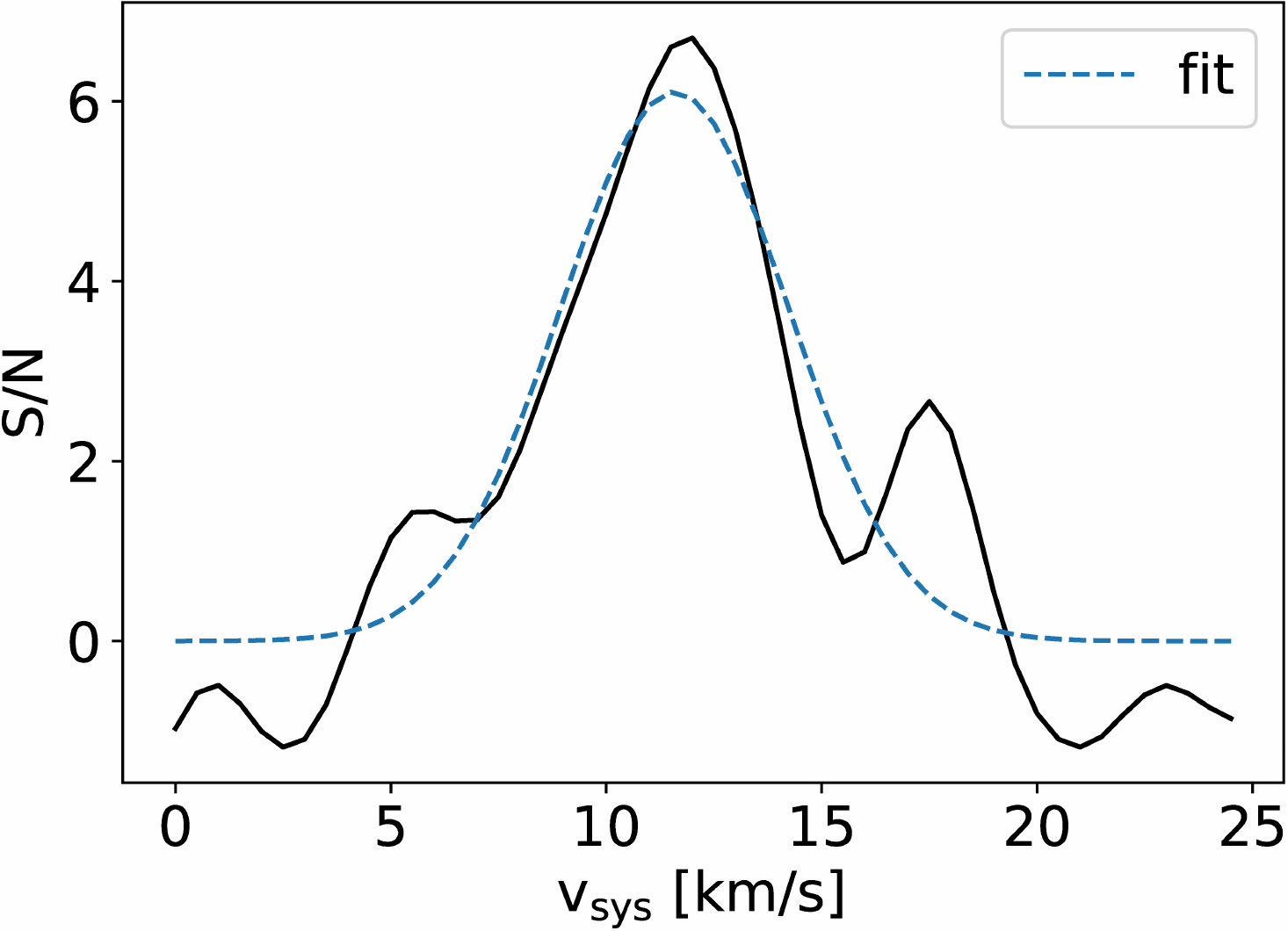}\\
    \includegraphics[width=.3\linewidth]{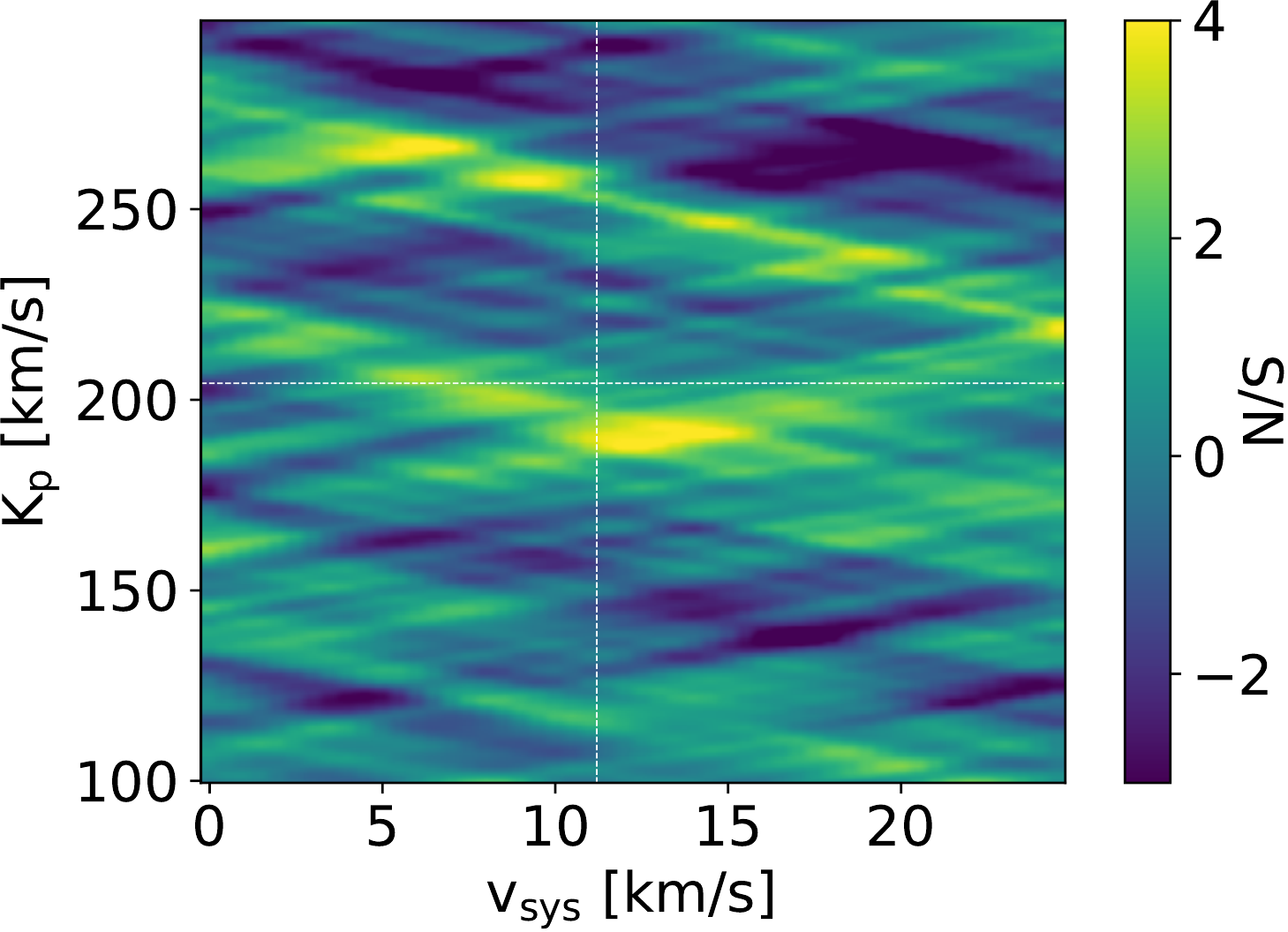}\hspace{.5cm}
    \includegraphics[width=.3\linewidth]{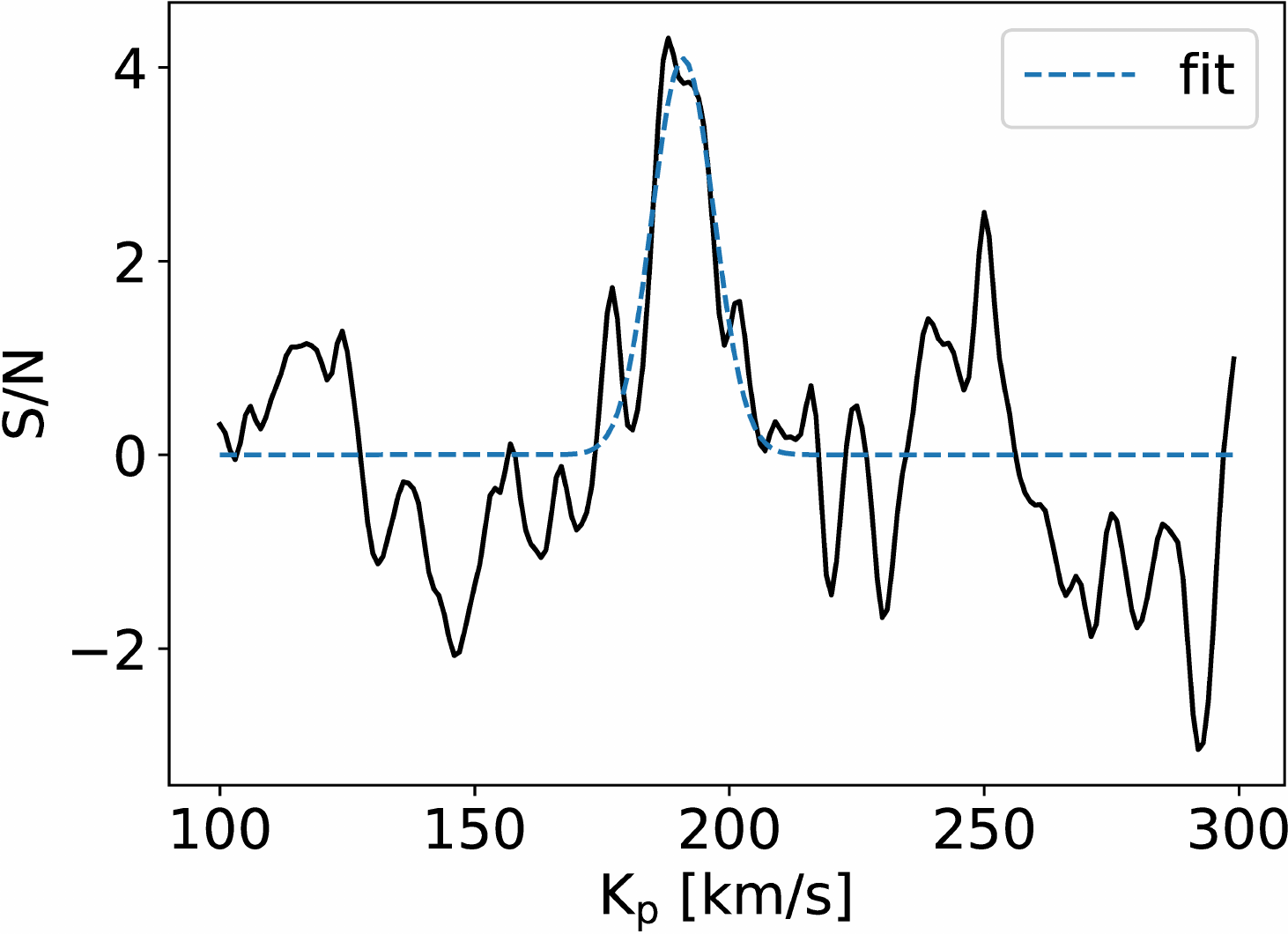}\hspace{.5cm}
    \includegraphics[width=.3\linewidth]{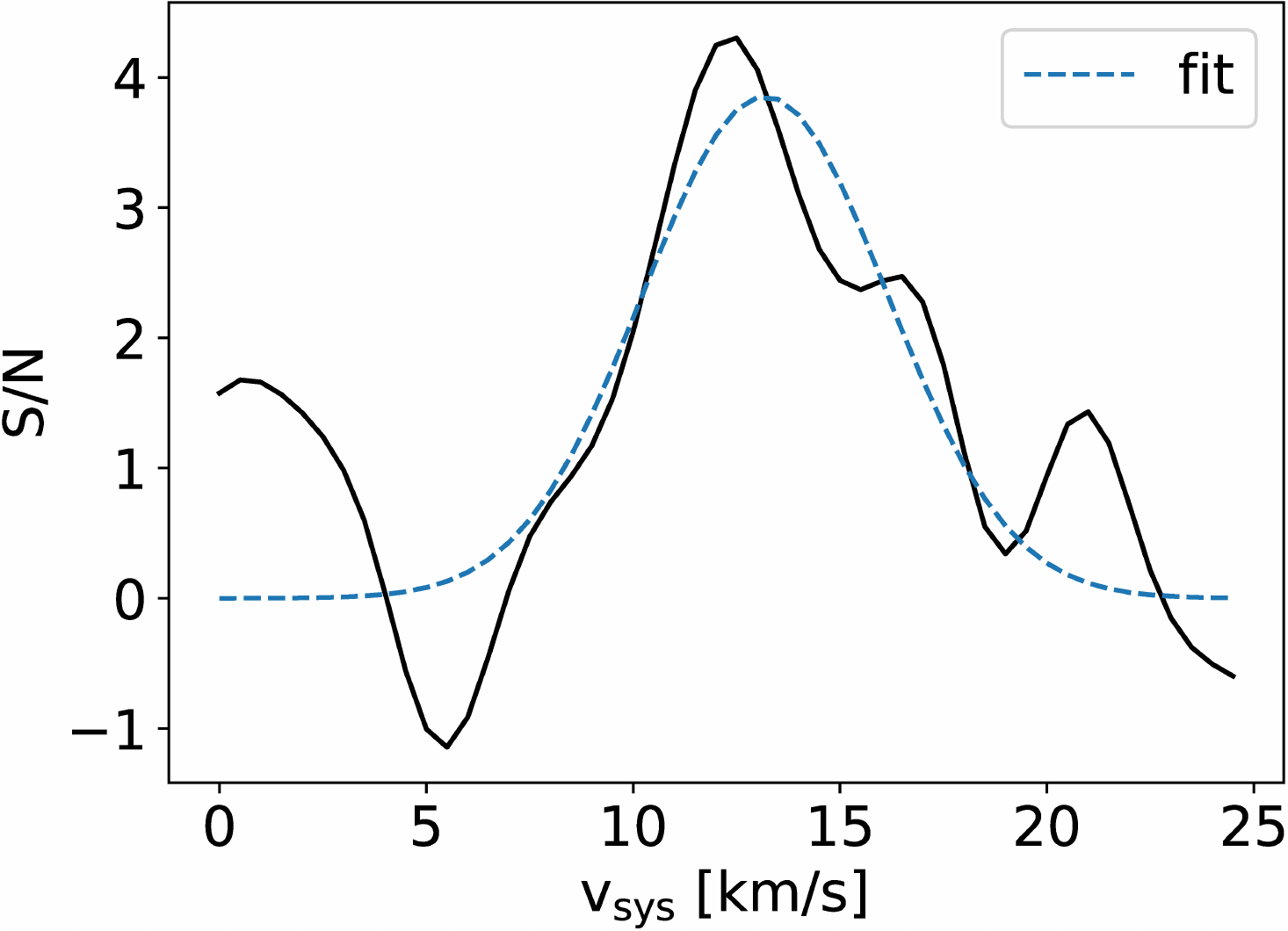}\\
    \includegraphics[width=.3\linewidth]{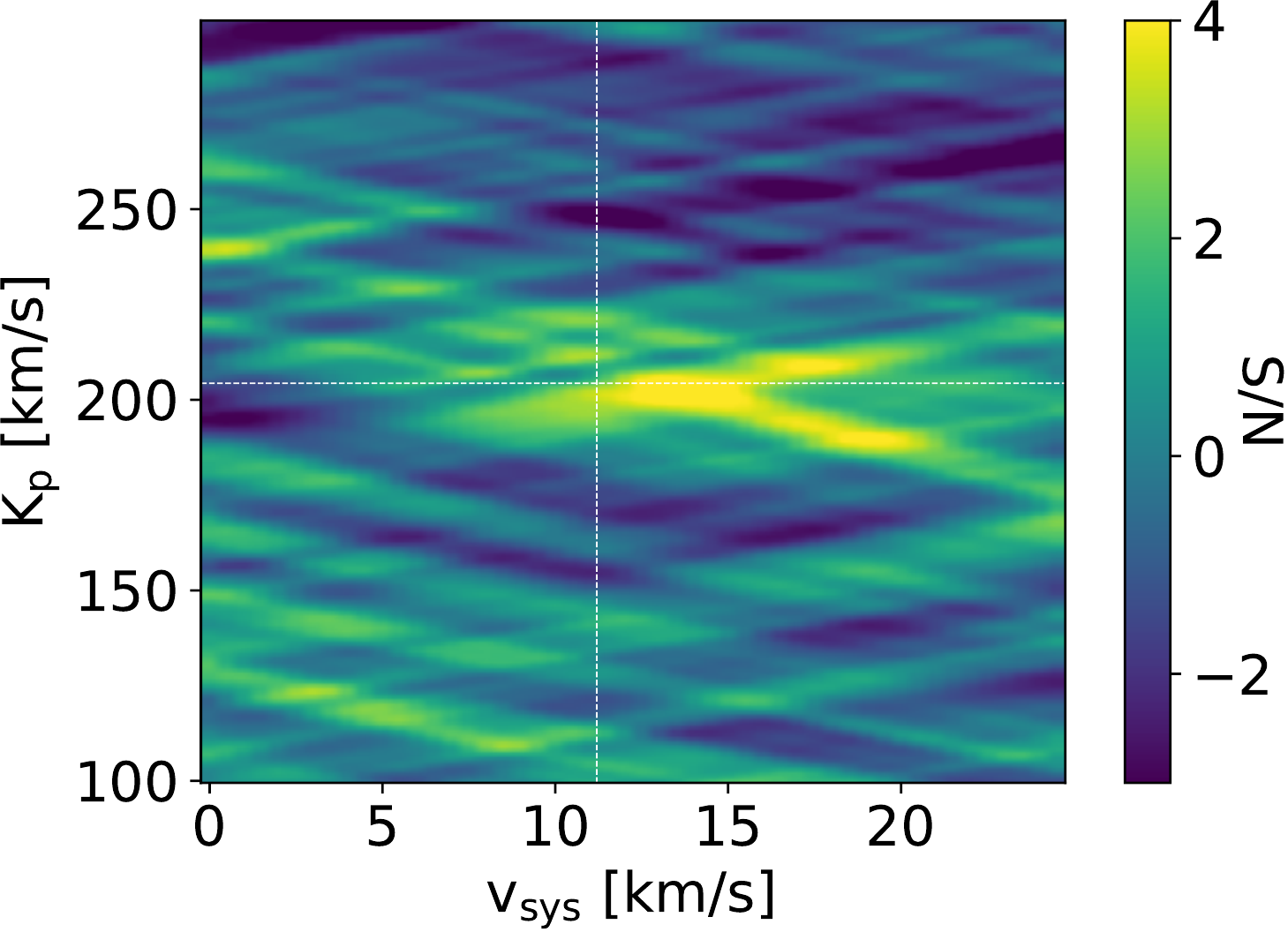}\hspace{.5cm}
    \includegraphics[width=.3\linewidth]{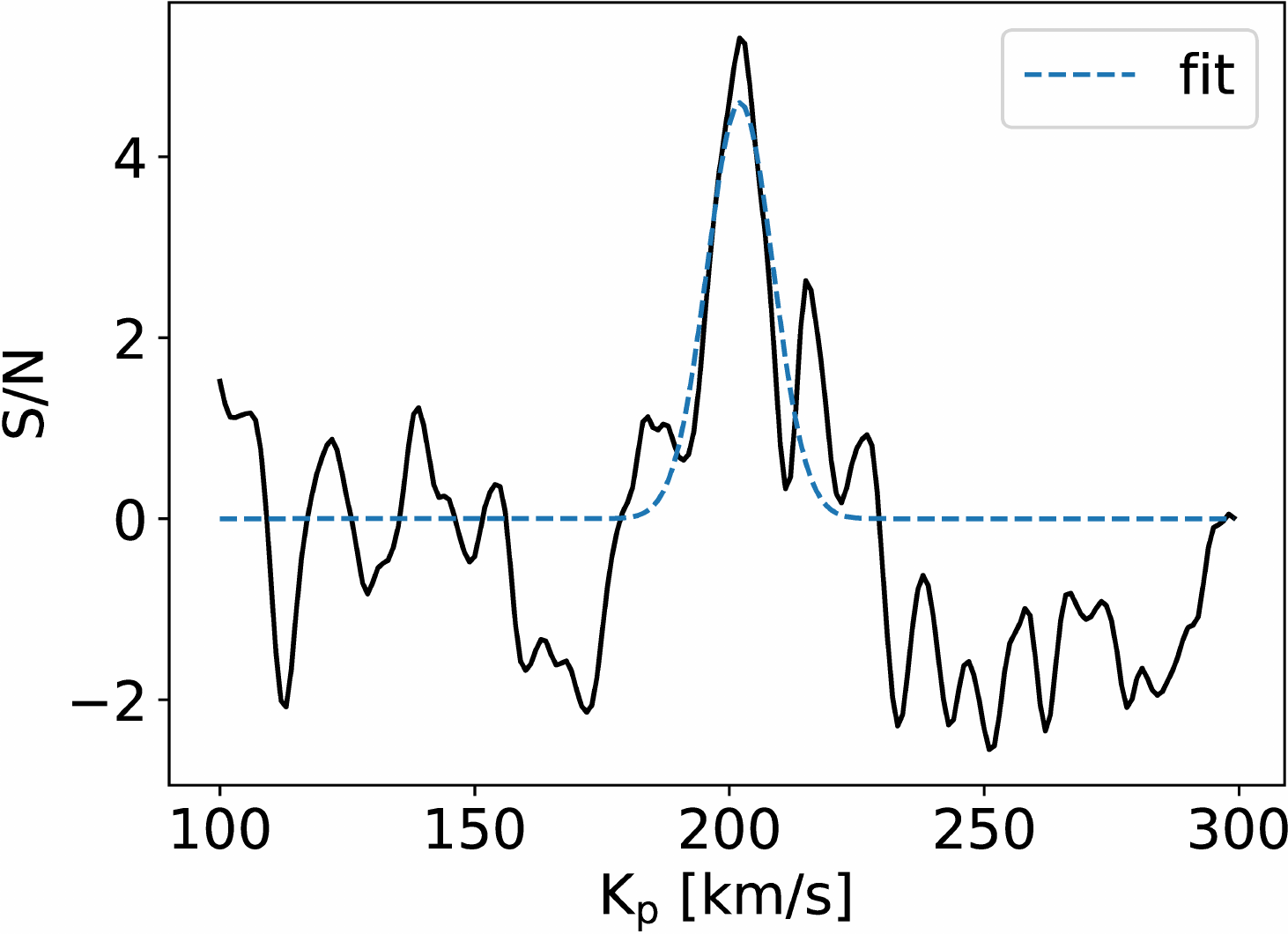}\hspace{.5cm}
    \includegraphics[width=.3\linewidth]{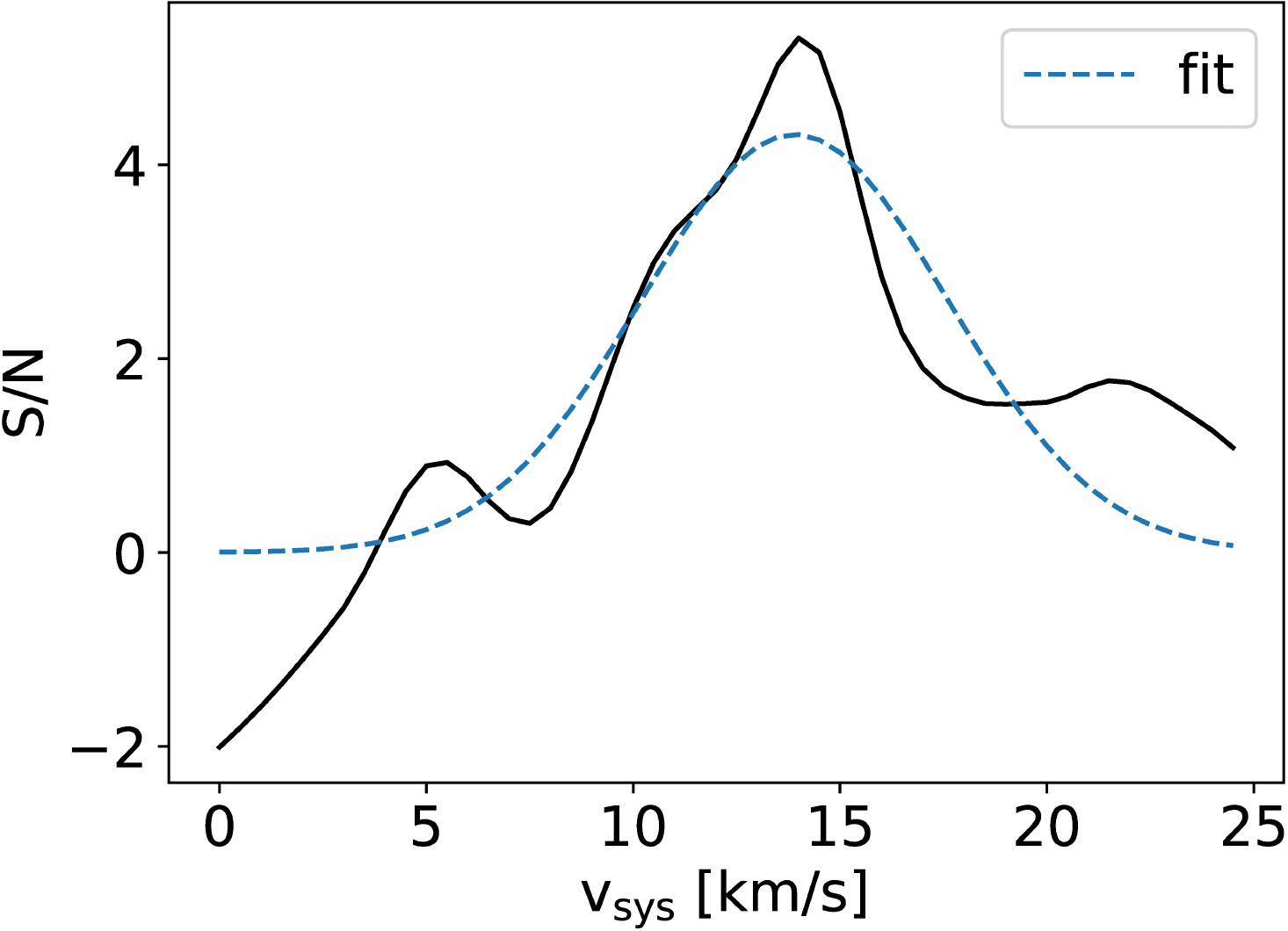}    
    \caption{Atmospheric planetary signal obtained with the \ion{Fe}{i}, \ion{Cr}{i} and \ion{Ti}{i} templates (top, central and bottom row respectively). The left column shows the \kp--\vsys\ bi-dimensional map, with the color bar encoding the significance of the detection. The cross in white dashes marks the position where the planetary signal expected according to Table~\ref{tab:parameters}. The central (right) column shows the cut along the vertical (horizontal) direction in the map and corresponding to the peak detection, together with its gaussian best fit (blue dashed line).}
    \label{fig:kpvsysFeI}
\end{figure*}

In Fig.~\ref{fig:kpvsysFeI} we show the \kp--\vsys\ maps for \ion{Fe}{I}, \ion{Cr}{I} and \ion{Ti}{I}: while they ubiquitously show positive and negative artifacts, it is striking that they also show a strong positive signal close to where we expect the planetary emission signal (see Table~\ref{tab:parameters}). We thus claim clear detection with a S/N larger than 4 (Table~\ref{tab:peaks}). For the other templates we did not find anything statistically significant, meaning that those species do not produce a \ac{CCF} signal stronger than noise.

For each map, we extracted the horizontal slice intersecting the peak. To quantify the center of the detected emission feature and the corresponding uncertainty, we computed a Gaussian fit to estimate \vsys. Similarly, we operated a vertical slice through the map peak and fit a Gaussian function to estimate \kp\  (see central and right columns in Fig.~\ref{fig:kpvsysFeI}). In both cases, due to correlated noise introduced by the CCF technique, it is difficult to estimate the uncertainties on the centroids. For simplicity we thus used the widths of the gaussian fits as the uncertainties for the best fit parameters. The \vsys\ and \kp\ best estimates are listed together with their corresponding uncertainties in Table~\ref{tab:peaks}. We found that the measurements all agree within uncertainties among each other and with the estimates listed in Table~\ref{tab:parameters}.

\begin{table}
\caption{Summary of the detected species in the \kp--\vsys\ maps.}\label{tab:peaks}      
\centering          
\begin{tabular}{l c c c}     
\hline\hline       
Element & \kp~[km/s] & \vsys~[km/s] & S/N\\
\hline
\ion{Fe}{i} & 201$\pm$5 & 12$\pm$3 & 6.7\\
\ion{Cr}{i} & 191$\pm$6 & 13$\pm$3 & 4.3\\
\ion{Ti}{i} & 202$\pm$4 & 14$\pm$4 & 5.3\\
\hline                  
\end{tabular}
\end{table}

\section{Conclusions}\label{sec:conclusions}

In this work we confirm the atmospheric signal for \mascb, an \ac{UHJ}. We observed its dayside at two epochs, before and after the secondary eclipse respectively, using the PEPSI high-resolution spectrograph mounted at the LBT. By means of cross-correlation techniques we detected for the first time \ion{Ti}{i}, \ion{Cr}{i} and \ion{Fe}{i} emission in the atmosphere of the planet. This detection confirms that the atmosphere of \mascb\ can be detected through emission spectroscopy \citep{Holmberg2022}, despite the fact that high-resolution transmission spectroscopy has been unsuccessful so far \citep[][]{Stangret2022,Casasayas2022}.

The system of reference of all our detections is fully consistent with the expected systemic velocity and radial velocity amplitude of \mascb, thus confirming their planetary nature (Table \ref{tab:peaks}). The absence of a signal in transmission in previous publications is thus likely due to the high gravity of the planet or to the overlap of the planetary track with the Doppler shadow as discussed by \citet{Casasayas2022}. Another interesting interpretation points to the 3D nature of \mascb's atmosphere, with atomic species more visible in the planetary dayside because of being dissociated due to the higher temperatures with respect to the terminator.

In a thermal radiation spectrum the flux of the spectral lines originates from higher altitudes than the continuum \citep[][]{Yan2022}. The detection of lines in emission from the planet thus indicates hotter temperatures at higher altitudes, and confirms for \mascb\ the thermal inversion already observed by \citet{Holmberg2022} and observed in the atmosphere of other \acp{UHJ} \citep[e.g., ][]{Lothringer2019,Pino2020,nugroho2020,Yan2022,borsa2022}. For these planets, the thermal inversion is caused by the high UV flux received from the host star (which is an A-type in most cases).

The detection of Ti in the atmosphere of MASCARA-1b is of particular relevance. Recently, by analyzing high-resolution emission spectroscopy of the UHJ WASP-121~b and not finding traces of Ti, \citet{Hoeijmakers2022} proposed the possibility of Ti cold trapping. In particular, they propose a sharp transition above which Ti become detectable in exoplanetary atmospheres to be between the $\rm T_{eq}$ of WASP-121~b (2358$\pm$52K) and WASP-189~b (2641$\pm$34K) \citep{prinoth2022}. In this work, we find evidence of the presence of Ti in the atmosphere of \mascb, with a $T_{\rm eq}$ of $2570^{+50}_{-30}$ \citep{Talens2017}, which further reduces the transition temperature interval.

High-resolution emission spectroscopy is confirmed to be an excellent and complementary technique to transmission spectroscopy when investigating the atmosphere of \acp{UHJ}. This is the first case where an absence of signal in transmission has been subsequently followed by a detection in emission, as it was suggested for this planet by \citet[][]{Casasayas2022}.

\begin{acknowledgements}

The LBT is an international collaboration among institutions in the United States, Italy and Germany. LBT Corporation Members are: The University of Arizona on behalf of the Arizona Board of Regents; Istituto Nazionale di Astrofisica, Italy; LBT Beteiligungsgesellschaft, Germany, representing the Max-Planck Society, The Leibniz Institute for Astrophysics Potsdam, and Heidelberg University; The Ohio State University, representing OSU, University of Notre Dame, University of Minnesota and University of Virginia.

GSc acknowledges support from CHEOPS ASI-INAF agreement n. 2019-29-HH.0. GSc is grateful to C.M.C.S. for the best funds coming in small packages (agreement Aug,30 2019).

We acknowledge the support by INAF/Frontiera through the "Progetti Premiali" funding scheme of the Italian Ministry of Education, University, and Research and from PRIN INAF 2019

This research was supported by the Excellence Cluster ORIGINS which is funded by the Deutsche Forschungsgemeinschaft (DFG, German Research Foundation) under Germany's Excellence Strategy - EXC-2094 - 390783311.

KP acknowledges funding from the German Leibniz-Gemeinschaft under project number P67/2018.

\end{acknowledgements}

\bibliographystyle{aa}
\bibliography{references}



\end{document}